\documentclass[9pt,twocolumn,twoside]{opticajnl}
\journal{opticajournal} 

\setboolean{shortarticle}{true}

\usepackage{lineno}
\usepackage{braket}
\usepackage{soul, xcolor}
\title{Quantitative Lineshape Analysis for Arbitrary Inhomogeneity in Two-Dimensional Coherent Spectroscopy}

\author[1]{Bhaskar De}
\author[1]{Pradeep Kumar}
\author[1]{Krishna K. Maurya}
\author[1]{Rishabh Tripathi}
\author[1,*]{Rohan Singh}

\affil[1]{Department of Physics, Indian Institute of Science Education and Research Bhopal, Bhopal 462066, India}

\affil[*]{rohan@iiserb.ac.in}

\begin{abstract}
Two-dimensional coherent spectroscopy (2DCS) provides simultaneous measurement of homogeneous and inhomogeneous linewidths through quantitative lineshape analysis.
However, conventional lineshape analysis methods assume Gaussian inhomogeneity, limiting its applicability to systems with non-Gaussian inhomogeneity.
We present a quantitative lineshape analysis method incorporating arbitrary inhomogeneity using a bivariate spectral distribution function in 2DCS simulations. 
An algorithm is developed to extract the homogeneous linewidth and arbitrary inhomogeneous distribution from experimentally-measured 2D spectrum.
We demonstrate this framework for a quantum-well-exciton resonance with non-Gaussian inhomogeneity.
This work broadens the scope of quantitative lineshape analysis for studying materials with non-Gaussian inhomogeneity.
\end{abstract}

\setboolean{displaycopyright}{false} 

\begin{document}

\setstcolor{red}

\maketitle

Homogeneous and inhomogeneous broadening manifest similarly as spectral broadening \cite{Mukamel1995,Klingshirn2012} despite having distinct origins.
Thus, conventional linear spectroscopy cannot separate them \cite{Moody2015}. 
Accurate quantification of both homogeneous and inhomogeneous linewidths is essential for material characterization and coherent-control applications. 
Nonlinear spectroscopic techniques such as two-dimensional coherent spectroscopy (2DCS) can decipher their effects by resolving the spectrum into two frequency axes \cite{Cho2008,Smallwood2018} and quantify them through lineshape analysis \cite{Siemens2010,Bell2015}.
Moreover, lineshape analysis has proven to be useful in quantifying spectral diffusion through frequency-frequency correlation function (FFCF) \cite{Roberts2006}. 

Traditional lineshape analysis methods for 2D spectra are developed assuming Gaussian inhomogeneity \cite{Roberts2006,Lazonder2006,Siemens2010,Bell2015,Kumar2024} following the central limit theorem \cite{Mukamel1995}. 
However, not all materials necessarily exhibit Gaussian distribution. Experimentally, continuous non-Gaussian inhomogeneity has been observed in various systems \cite{Schnabel1992,Cho2008,Kasprzak2010, Moody2011,Martin2018}.
Gaussian-inhomogeneity-based lineshape analysis may be inaccurate in such cases.
On the other hand, non-Gaussian inhomogenity are often modeled using specific assumptions about frequency fluctuations over ensemble or time \cite{Cho2008,Hoshino2025}.
For both cases, the assumptions vary between systems depending on their interactions with the local environment \cite{Roy2011}.
As a result, these methods are often system-specific and the quantified inhomogeneity is inherently limited by the underlying assumptions.
A generalized deconvolution method was used to extract the Green's function from 2D spectra \cite{Richter2018}.
Nevertheless, a simple and flexible lineshape analysis technique incorporating arbitrary inhomogeneity has not been reported to best of our knowledge. 

In this work, we present a simulation and lineshape analysis technique for 2DCS for arbitrary inhomogeneity.
Our method describes the third-order nonlinear response as a convolution of the homogeneous response with a bivariate spectral distribution function (BSDF).
Therefore, we establish a direct correspondence between 2D spectrum and the arbitrary inhomogeneity.
This relation enables us to extract arbitrary inhomogeneous distributions without predefined assumptions.
Furthermore, we define a weighted correlation of BSDF to quantify spectral diffusion in non-Gaussian inhomogeneous systems.
We apply this technique to the quantum-well (QW) excitonic resonance in a GaAs interfacial quantum dot (IQD) sample, to illustrate the fitting procedure and its ability to reveal the arbitrary inhomogeneous distribution from experimental 2D spectrum.
This method advances the quantitative lineshape analysis of 2DCS beyond Gaussian-inhomogeneity-based models.

\begin{figure}[t]
\centering\includegraphics{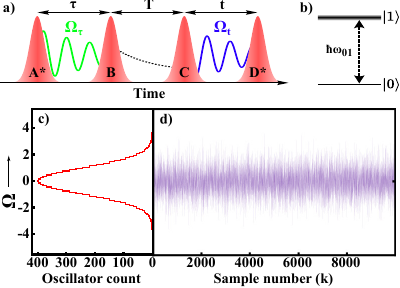}
\caption{(a) Pulse sequence for rephasing 2DCS experiment: three excitation pulses A*, B and C are applied with inter-pulse delays $\tau$ and $T$, respectively.
Emission occurs during delay $t$ and D* is the detection pulse.
During delays $\tau$ and $t$, the signal phase evolves at frequencies $\Omega_{\tau}$ and $\Omega_{t}$, respectively.
(b) A two-level system with resonance energy $\hbar\omega_{01}$.
(c) A discrete Gaussian inhomogeneous oscillator distribution, $g_\text{1D}(\Omega)$
(d) One realization of the stochastic variable generated from the oscillator distribution in (c) using histogram-based sampling.}
\label{fig:Pic1}
\end{figure}

In a typical 2DCS experiment, three excitation pulses with inter-pulse delays $\tau$ and $T$ excite the samples.
The radiated four-wave mixing (FWM) signal can be detected by heterodyning with a fourth pulse arriving after a delay $t$ after the excitation pulses.
The excitation pulses are schematically shown in Fig. \ref{fig:Pic1}(a).
The generalized time-domain rephasing FWM signal is of the form
\begin{equation}
\label{eq:eqn2}
    S_{h}(\tau,T,t,\Omega_{\tau},\Omega_{t}) \propto e^{i\left(\Omega_{\tau}\tau - \Omega_T T -\Omega_{t}t\right)} e^{-(\gamma_{\tau}\tau +\gamma_T T+\gamma_{t}t)}\Theta(\tau,T,t).
\end{equation}
Here, $\Omega_j$ and $\gamma_j$ indicate the phase evolution and decay during the various delays, which are indicated by the subscripts.
$\Theta$ is the Heaviside step function. 
The above analytical solution is obtained by a perturbative solution of the relevant optical Bloch equations (OBEs) assuming excitation with $\delta$-function pulses in time.
The variables $\Omega_{\tau}$ and $\Omega_{t}$ indicate the position of the peak in a 2D spectra, which is obtained by taking a Fourier transform of the time-domain signal $S_h$ along delays $\tau$ and $t$.
For the case of a two-level system (2LS) shown in Fig. \ref{fig:Pic1}(b), we can use \eqref{eq:eqn2} to obtain the correct rephasing signal for $\Omega_{\tau}=\Omega_t=\omega_{01}$ the transition energy for the 2LS, $\Omega_T=0$, $\gamma_{\tau}=\gamma_t=\gamma_{01}$ the dephasing rate of the transition and $\gamma_{T}=\Gamma_1$ the population decay rate of the excited state. 

At this point, inhomogeneity is generally assumed to be Gaussian and multiplied with the homogeneous response to obtain the inhomogeneous response. However, we take a different approach to replicate ensemble-averaging and inhomogeneity. We consider a discrete inhomogeneous distribution of transition frequencies, $g_{1D}(\Omega)$, as shown in Fig. \ref{fig:Pic1}(c) for $N=10^4$ oscillators; we have shifted the center frequency to 0 for simplifying the calculations.
This distribution represents the statistical spread of exciton energies due to ensemble averaging.
To model this inhomogeneity, we treat the transition frequencies as stochastic variables, where each oscillator has a transition frequency $\Omega^{[k]}$ drawn from the distribution, i.e., $\Omega^{[k]} \sim g_{1D}(\Omega)$.
Here, and in rest of this work, we indicate the $k$-th element of a discrete variable with the superscript $[k]$.
An example realization of these randomly sampled frequencies is shown in Fig. \ref{fig:Pic1}(d), illustrating the fluctuations that arise due to the statistical nature of the distribution.

We define a BSDF $g_{2D}(\Omega_\tau, \Omega_t, M)$ using this stochastic variable.
The two frequency variables correspond to the two frequency axes in a 2D spectrum and $M$ denotes the normalized FFCF.
In general, inhomogeneous broadening can result in unique phase-evolution frequencies during delays $\tau$ and $t$.
The BSDF naturally captures this phenomena by accounting for the fact that two independent samples drawn from the same distribution $g_{1D}(\Omega)$ can differ, with their relation determined by the correlation $M$.

Assuming that the homogeneous response is uniform across the inhomogeneous distribution, the total time-domain signal for the inhomogeneous distribution can be obtained by a convolution of the homogeneous response with the BSDF
\begin{eqnarray}
    S(\tau,T,t) = \sum_{i,j=1}^{N}S_{h}(\tau,T,t,\Omega^{[i]}_{\tau},\Omega^{[j]}_{t}) g_{\text{2D}}(\Omega^{[i]}_{\tau},\Omega^{[j]}_{t},M).
    \label{eq:eqn4}
\end{eqnarray}
$g_{2D}$ takes the form of a two-dimensional elliptical Gaussian in case of a Gaussian inhomogeneous distribution \cite{Cundiff1994,Liu2021}.
The convolution can be performed analytically to obtain
\begin{eqnarray}
\label{eq:eqn1}
    S(\tau,T,t) = S_{h}(\tau,T,t,\omega_{01},\omega_{01}) e^{-\frac{1}{2}\sigma^2(\tau^2-2M\tau t+t^2)}.
\end{eqnarray}
for a Gaussian inhomogeneous distribution centered at $\omega_{01}$\cite{Roberts2006, Singh2016}.
The Fourier transform $\tilde{S}(\omega_\tau,T,\omega_t)$ captures the homogeneous and inhomogeneous broadening along the diagonal and cross-diagonal directions, respectively, of the rephasing 2D spectrum \cite{Siemens2010}.

It is not possible to find a simple analytical solution of the inhomogeneous response analogous to \eqref{eq:eqn1} for an arbitrary inhomogeneity.
In this case, we define a discrete BSDF, in order to simulate the corresponding 2D spectra, as
\begin{eqnarray}
\label{eq:eqn6}
g_{\text{2D}}(M) = [g_{\text{1D}}(\Omega_\tau) \otimes g_{\text{1D}}(\Omega_t) P_{\text{D}}(\Omega_\tau - \Omega_t,a;M)]^{\frac{1}{1+M}}.
\end{eqnarray}
The frequency variables of $g_{2D}$ in \eqref{eq:eqn6} are omitted for brevity.
The outer product defines $g_{2D}$ in case of no correlation between the $\Omega_\tau$ and $\Omega_t$ axes, i.e., $M=0$.
$P_D$ is a probability distribution with spread-control parameter $a$ that is used to narrow the uncorrelated $g_{2D}(M=0)$ for $M>0$.
We note that \eqref{eq:eqn6} can be derived for a Gaussian inhomogeneity (see Section 1 of Supplement 1); we assume that the same relation holds even for arbitrary inhomogeneity.


We define a weighted correlation coefficient
\begin{eqnarray}
\label{eq:eqn5}
r_w = \frac{\sum\limits_{i,j=1}^N g_{\text{2D}}^{[i,j]} (\Omega^{[i]}_{\tau} - \overline{\Omega}_{\tau}) (\Omega^{[j]}_{t} - \overline{\Omega}_t)}{\sum\limits_{i=1}^N g_{\text{1D}}^{[i]}(\Omega^{[i]} - \overline{\Omega})^2}.
\end{eqnarray}
For a Gaussian inhomogeneity, $r_w$ is the same as $M$ (see Section 2 of Supplement 1).
We use an iterative process to modify $P_D$ such that the calculated value of $r_w$ is equal to the desired value of $M$.

\begin{figure}[t]
\centering\includegraphics{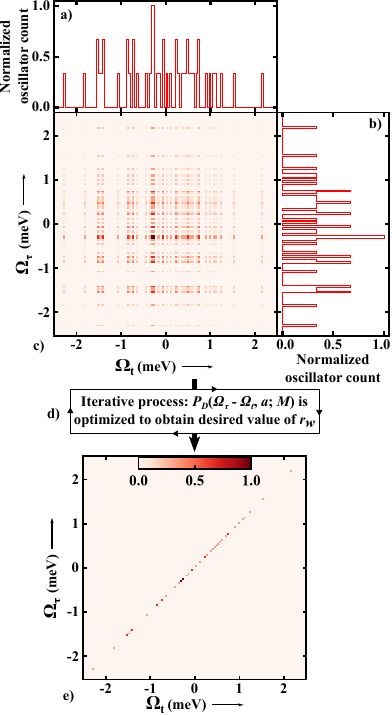}
\caption{(a) and (b) Arbitrary inhomogeneous oscillator distribution. (c) Uncorrelated Bivariate spectral distribution function. (d) Iterative process to obtain desired value of $r_w$ by optimizing the value of $a$ using \eqref{eq:eqn6} and \eqref{eq:eqn5}. (e) BSDF with correlation $r_w = 1$.}
\label{fig:Pic2}
\end{figure}

We now illustrate the procedure to simulate a rephasing 2D spectrum based on the formalism discussed above.
We select a collection of 50 oscillators from the distribution of $10^4$ oscillators that form the inhomogeneous distribution in Fig. \ref{fig:Pic1}(c).
The inhomogeneous distribution for these 50 oscillators is shown in Figs. \ref{fig:Pic2}(a) and \ref{fig:Pic2}(b), along the $\Omega_\tau$ and $\Omega_t$ axes, respectively.
The corresponding uncorrelated BSDF is shown in Fig. \ref{fig:Pic2}(c).
This uncorrelated spectral map is narrowed through a symmetric peak function $P_D$; we use a Gaussian with standard deviation $a$.
We calculate the BSDF using \eqref{eq:eqn6} using an initial value of $a$.
The correlation $r_w$ for the narrower spectral map is calculated using \eqref{eq:eqn5} and compared to the desired value of $M$.
The value of $a$ is varied till we obtain $r_w = M$, as indicated in Fig. \ref{fig:Pic2}(d).
The BSDF for $M=1$ obtained through this procedure is shown in Fig. \ref{fig:Pic2}(e).
As expected, the sum of the BSDF along either $\Omega_\tau$ or $\Omega_t$ axes gives the inhomogeneous distribution.

\begin{figure}[t]
\centering\includegraphics{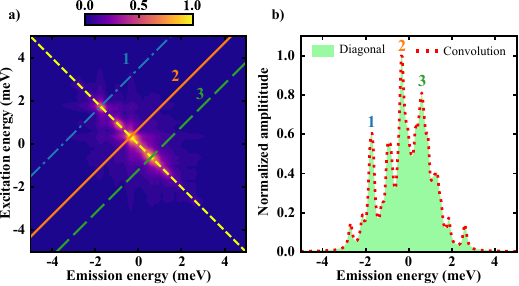}
\caption{(a) Simulated rephasing 2D spectrum for the oscillator distribution shown in  Fig. \ref{fig:Pic2}(a).
(b) Comparison of the diagonal slice of the simulated 2D spectrum (green shaded region) and the convolution of Lorentzian with $g_\text{1D}(\Omega)$ (red dashed line).
The labels $1-3$ in (b) indicate the frequencies for which the cross-diagonal slices in (a) are taken.}
\label{fig:Pic3}
\end{figure}


Subsequently, we simulated the 2D spectrum for $M = 1$ and $\gamma_{01} = 0.1 \, \text{meV}$ using \eqref{eq:eqn4}, which is shown in Fig. \ref{fig:Pic3}(a).
The diagonal slice of this spectrum is represented by the green shaded region in Fig. \ref{fig:Pic3}(b).
As expected, the diagonal slice shows several discete peaks rather than a single, smooth peak obtained for Gaussian inhomogeneity using \eqref{eq:eqn1}.
Additionally, the convolution of a Lorentzian having half-width at half maximum $\gamma_{01}$ with $g_\text{1D}(\Omega)$ is plotted as the red dashed line.
Interestingly, there is an excellent agreement between these two plots for arbitrary inhomogenity, which suggests that 
\begin{eqnarray}
\label{eq:eqn7}
\tilde{S}_{Diag}(\omega_{\tau'}) = g_\text{1D}(\omega_{\tau'})*\frac{1}{{\gamma_{01}}^2+{\omega_{\tau'}}^2}.
\end{eqnarray}
Here, $\tau'= t-\tau$ and $\omega_{\tau'}$ is Fourier transform with respect to $\tau'$.
\eqref{eq:eqn7} is equivalent to the Voigt profile of the diagonal slice obtained for Gaussian inhomogeneity \cite{Siemens2010}.

\begin{figure}[t]
\centering\includegraphics{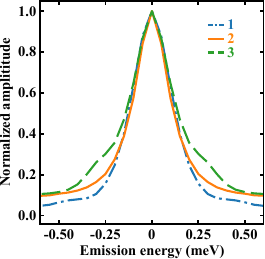}
\caption{Normalized amplitudes for cross-diagonal slices $1-3$ indicated in Fig. \ref{fig:Pic3}(a).
These slices show different lineshapes even for identical dephasing rate $\gamma_{01}$.}
\label{fig:Pic3_1}
\end{figure}

Figure \ref{fig:Pic3_1} shows several cross-diagonal slices $1-3$ taken across the simulated 2D spectrum in Fig. \ref{fig:Pic3}(a).
Although $\gamma_{01}$ was constant across the inhomogeneous distribution, the cross-diagonal slices show significant variations.
This observation suggests that the crossdiagonal lineshape is significantly affected by the number of oscillators in the spectral vicinity of a particular frequency.
Consequently, relying solely on the crossdiagonal width is insufficient for estimating the homogeneous linewidth.
Therefore, for arbitrary inhomogeneity, the entire 2D spectrum must be fitted to simultaneously quantify both the inhomogeneous distribution and $\gamma_{01}$.
This is a crucial consideration when studying spatially inhomogeneous samples, particularly with small excitation spots \cite{Martin2018}.

\begin{figure}[t]
\centering\includegraphics{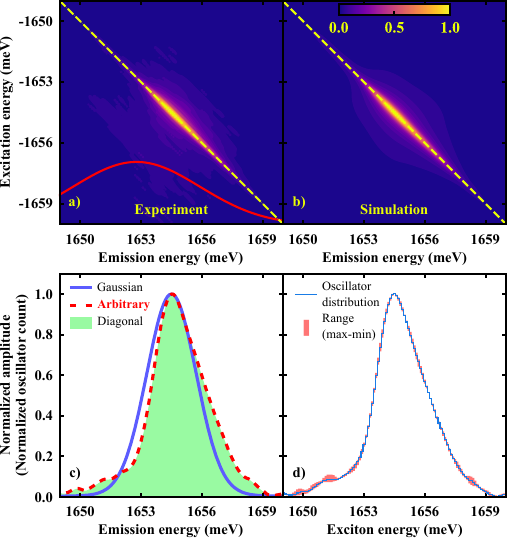}
\caption{(a) Experimental 2D rephasing spectrum showing the QW-exciton resonance in a GaAs IQD at 5 K. The red line shows the excitation-pulse spectrum. (b) Simulated rephasing 2D spectrum generated using the best-fit oscillator distribution and homogeneous linewidth ($\gamma_{01}$). (c) Comparison of the diagonal slices. Green shaded region represent the slice of experimental data. Blue line and red dashed line represent the slices from simulations assuming Gaussian inhomogeneity and arbitrary inhomogeneity, respectively.
(d) The extracted oscillator distribution shown in blue, reveals a non-Gaussian inhomogeneous distribution. The magenta shaded region shows the range of normalized oscillator count.}
\label{fig:Pic4}
\end{figure}

We leverage the simulation framework to simultaneously quantify the oscillator distribution and homogeneous linewidth from experimental data.
Figure \ref{fig:Pic4}(a) shows a 2D spectrum for the QW-exciton resonance in a 4.2-nm-wide GaAs IQD sample \cite{Moody2011}.
The experiment was performed with collinear excitation pulses, as shown in Fig. \ref{fig:Pic1}(a).
The excitation pulses (A*, B, and C) with repetition rate of 50 kHz and centered at 1653 meV are incident on the sample with an excitation power of 2 $\mu$W/beam.
Each excitation pulse is phase modulated and, consequently, tagged with a unique radio frequency using acousto-optic modulators (AOMs).
The sample temperature was maintained at 5 K.
The radiated FWM is heterodyned with the local oscillator (D*) and detected using a balanced detector.
Further details of the sample and experimental technique can be found elsewhere \cite{Tripathi2025}.

The diagonal slice shown as green shaded region in Fig. \ref{fig:Pic4}(c) is extracted from the 2D spectra in Fig. \ref{fig:Pic4}(a).
The diagonal slice clearly shows a non-Gaussian lineshape, which indicates a non-Gaussian inhomogeneous broadening.
Assuming an initial guess for $\gamma_{01}$, we begin the fitting procedure by finding the inhomogeneous distribution $g_\text{1D}(\Omega)$ such that the diagonal slice of the experimental 2D spectra satisfies \eqref{eq:eqn7}.
This $g_\text{1D}(\Omega)$ is used to generate $g_\text{2D}(M=1)$ following the procedure depicted in Fig. \ref{fig:Pic2}, which, in turn, is used to simulate a rephasing 2D spectrum.
Next, the entire 2D spectrum is fit by changing $\gamma_{01}$ while keeping $g_\text{1D}(\Omega)$ fixed.
The above iterative process is repeated till the simulated spectra closely matches the experimental 2D spectrum and the best-fit values for $g_\text{1D}(\Omega)$ and $\gamma_{01}$ are obtained.
The best-fit 2D spectrum generated using this algorithm is shown in Fig. \ref{fig:Pic4}(b), which closely matches the experimental data in Fig. \ref{fig:Pic4}(a) ($R^2 = 0.95$).
The dashed red line in Fig. \ref{fig:Pic4}(c) shows the diagonal slice from the simulated data.
We obtain an excellent agreement between the diagonal $(R^2 > 0.99)$ and cross diagonal $(R^2 = 0.97)$ (see Section 3 of Supplement 1) slices obtained from the measured and simulated 2D spectra.
As expected, fitting the 2D spectrum to analytical expressions obtained for Gaussian inhomogeneity given by \eqref{eq:eqn1} does not yield a good fit as indicated by the blue line.
We plot the measured inhomogeneous distribution in Fig. \ref{fig:Pic4}(d). 
The finite bandwidth of excitation spectrum may underestimate the oscillator distribution at spectrum edges \cite{Smallwood2017, Halaoui2024}. 
The homogeneous linewidth $\gamma_{01}$ is measured to be $0.0595 \pm 0.0004$ meV.
The uncertainties are the standard deviation measured by repeating the experiment four times.

In conclusion, this work introduces a simulation and fitting framework for 2DCS, enabling the simultaneous extraction of homogeneous linewidth and arbitrary inhomogeneous distributions.
This method extends, for the first time, the techniques developed for Gaussian inhomogeneity \cite{Siemens2010, Bell2015, Singh2016} to arbitrary inhomogeneity. This generalized approach for lineshape analysis is particularly important in cases where non-Gaussian inhomogeneity is significant \cite{Cho2008, Martin2018}.
Additionally, the weighted correlation coefficient $r_w$ provides a pathway to quantify and model spectral diffusion in the context of arbitrary inhomogeneity.

Finally, we note that we have demonstrated the fitting procedure for rephasing 2D spectrum assuming a uniform exponential coherence decay for all oscillators.
While these approximations are commonly used, many-body effects \cite{Kumar2024} and non-Markovian dynamics \cite{Lorenz2005} can result in nonexponential decay dynamics.
Similarly, the dephasing rate can vary with the transition energy \cite{Moody2011}.
Furthermore, some experimental implementations can only measure the total 2D correlation spectrum \cite{Cho2008}.
These effects can be incorporated in the simulations by appropriately modifying the homogeneous response in \eqref{eq:eqn2}, albeit with significantly higher computational cost for fitting (see Section 4 of Supplement 1).
These challenges motivate further investigations to optimize the fitting procedure for such complex situations.

\begin{backmatter}
\bmsection{Funding} The authors acknowledge support from the Science and Engineering Research Board (SERB), New Delhi under Project No. CRG/2023/003263.

\bmsection{Acknowledgment} We thank Steven T. Cundiff, Daniel Gammon, and Allan S. Bracker for providing the interfacial quantum dot sample.
We thank Kshitij for fruitful discussions.
B. D. and K. K. M. acknowledge the Ministry of Education, Government of India for support from the Prime Minister’s Research Fellows (PMRF) Scheme.

\bmsection{Disclosures} The authors declare no conflicts of interest.

\bmsection{Data availability} Data underlying the results presented in this paper are not publicly available at this time but may be obtained from the authors upon reasonable request.





\bmsection{Supplemental document}
See Supplement 1 for supporting content.
\end{backmatter}




\bibliography{OL_arbinhomo}


\end{document}


\maketitle

\section{Relation between Uncorrelated and Correlated BSDF}
The two-dimensional elliptical Gaussian function \cite{Cundiff1994,Lomsadze2020} is defined in its most generalized form as:
\begin{equation}
\label{eq:eqn_supp1}
    g_{\text{2D}}(\Omega_\tau, \Omega_t, M) = \frac{1}{2\pi\sigma_\tau \sigma_t\sqrt{1-M^2}}exp\left[-\frac{\frac{(\Omega_\tau-\overline{\Omega}_\tau)^2}{\sigma_\tau^2}-2M\frac{(\Omega_\tau-\overline{\Omega}_\tau)(\Omega_t-\overline{\Omega}_t)}{\sigma_\tau \sigma_t}+\frac{(\Omega_t-\overline{\Omega}_t)^2}{\sigma_t^2}}{2(1-M^2)}\right]
\end{equation}
As inhomogeneity is same in both frequency axes, we consider $\sigma_\tau=\sigma_t=\sigma$. The \eqref{eq:eqn_supp1} becomes
\begin{equation}
\label{eq:eqn_supp2}
    g_{\text{2D}}(\Omega_\tau, \Omega_t, M) = \frac{1}{2\pi\sigma^2\sqrt{1-M^2}}exp\left[-\frac{(\Omega_\tau-\overline{\Omega}_\tau)^2-2M(\Omega_\tau-\overline{\Omega}_\tau)(\Omega_t-\overline{\Omega}_t)+(\Omega_t-\overline{\Omega}_t)^2}{2\sigma^2(1-M^2)}\right]
\end{equation}
From \eqref{eq:eqn_supp2} we have,
\begin{equation}
\label{eq:eqn_supp10}
    g_{\text{2D}}(\Omega_\tau, \Omega_t, M) = Aexp\left[-\frac{(\Omega_\tau-\overline{\Omega}_\tau)^2-2M(\Omega_\tau-\overline{\Omega}_\tau)(\Omega_t-\overline{\Omega}_t)+(\Omega_t-\overline{\Omega}_t)^2}{2\sigma^2(1-M^2)}\right]
\end{equation}

where $A =  \frac{1}{2\pi\sigma^2\sqrt{1-M^2}} $.
\begin{eqnarray}
\label{eq:eqn_supp12}
    g_{\text{2D}}(\Omega_\tau, \Omega_t, M) = \left[A'exp\left[-\frac{(\Omega_\tau-\overline{\Omega}_\tau)^2-2M(\Omega_\tau-\overline{\Omega}_\tau)(\Omega_t-\overline{\Omega}_t)+(\Omega_t-\overline{\Omega}_t)^2}{2\sigma^2(1-M)}\right]\right]^\frac{1}{1+M} \\ \nonumber= \left[A'exp\Bigg[-\frac{(\Omega_\tau-\overline{\Omega}_\tau)^2}{2\sigma^2}\Bigg]exp\Bigg[-\frac{(\Omega_t-\overline{\Omega}_t)^2}{2\sigma^2}\Bigg]exp\Bigg[-\frac{M(\Omega_\tau-\overline{\Omega}_\tau-(\Omega_t-\overline{\Omega}_t))^2}{2\sigma^2(1-M)}\Bigg]\right]^\frac{1}{1+M}
\end{eqnarray}
For simplification in calculation we shift the center to 0. Thus, we can consider $\overline{\Omega}_\tau = 0$ and $\overline{\Omega}_t = 0$. Therefore, \eqref{eq:eqn_supp12} reduces to
\begin{equation}
\label{eq:eqn_supp13}
    g_{\text{2D}}(\Omega_\tau, \Omega_t, M) = \left[A'exp\Bigg[-\frac{\Omega_\tau^2}{2\sigma^2}\Bigg]exp\Bigg[-\frac{\Omega_t^2}{2\sigma^2}\Bigg]exp\Bigg[-\frac{M(\Omega_\tau-\Omega_t)^2}{2\sigma^2(1-M)}\Bigg]\right]^\frac{1}{1+M}
\end{equation}
We can drop the constant $A'$ as we numerically normalized the $g_\text{2D}$ in our simulation and fitting. Therefore, we can write as
\begin{equation}
\label{eq:eqn_supp14}
    g_{\text{2D}}(M) = [g_{\text{1D}}(\Omega_\tau) \otimes g_{\text{1D}}(\Omega_t) P_{\text{D}}(\Omega_\tau - \Omega_t,a;M)]^{\frac{1}{1+M}}.
\end{equation}
where $g_\text{1D}(\Omega)$ is the oscillator distribution.
We are approximating the term $exp\Bigg[-\frac{M(\Omega_\tau-\Omega_t)^2}{2\sigma^2(1-M)}\Bigg]$ as $P_\text{D}$.
\section{Weighted Correlation}
The statistical correlation coefficient between two variable $\Omega_\tau$ and $\Omega_t$ is defined as 
\begin{equation}
\label{eq:corr_coeff}
    r_w = \frac{Cov(\Omega_\tau,\Omega_t)}{\sqrt{Var(\Omega_\tau)}\sqrt{Var(\Omega_t)}}
\end{equation}
As the 1D oscillator distribution for $\Omega_\tau$ and $\Omega_t$ is considered same, for discrete variable the \eqref{eq:corr_coeff} becomes  
\begin{equation}
\label{eq:eqn_supp6}
r_w = \frac{\sum\limits_{i,j=1}^N g_{\text{2D}}^{[i,j]} (\Omega^{[i]}_{\tau} - \overline{\Omega}_{\tau}) (\Omega^{[j]}_{t} - \overline{\Omega}_t)}{\sum\limits_{i=1}^N g_{\text{1D}}^{[i]}(\Omega^{[i]} - \overline{\Omega})^2}
\end{equation}
\begin{equation}
\label{eq:eqn_supp7}
    r_w = \frac{Cov(\Omega_\tau,\Omega_t)}{Var(\Omega)}
\end{equation}
\begin{equation}
\label{eq:eqn_supp8}
    r_w = \frac{\int\limits_{-\infty}^{\infty}\int\limits_{-\infty}^{\infty}(\Omega_\tau-\overline{\Omega}_\tau)(\Omega_t-\overline{\Omega}_t)g_\text{2D}(\Omega_\tau,\Omega_t;M)d\Omega_\tau d\Omega_t}{\int\limits_{-\infty}^{\infty}(\Omega-\overline{\Omega})^2g_\text{1D}(\Omega)d\Omega}
\end{equation}
Using \eqref{eq:eqn_supp1} and considering $\sigma_\tau = \sigma_t = \sigma$, we get
\begin{equation}
\label{eq:eqn_supp9}
    r_w = \frac{M\sigma_\tau\sigma_t}{\sigma^2}=M
\end{equation}

\section{Cross-diagonal slice}
\begin{figure}[htbp]
\centering\includegraphics{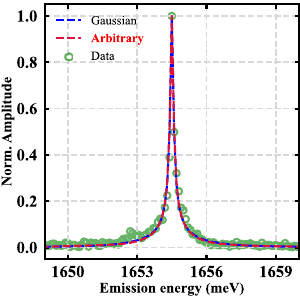}
\caption{The cross-diagonal slice of the experimental 2D spectrum shown with green circles. The blue and red dashed line shows the cross-diagonal slice of simulation from Gaussian and arbitrary inhomogeneity, respectively.}
\label{fig:cross_diagonal}
\end{figure}

The cross-diagonal slice of 2D spectrum is shown in Fig. \ref{fig:cross_diagonal}. The experimental 2D spectrum, and the diagonal slice has been shown in main text Figs. 5(a) and 5(c), respectively.
Both inhomogeneity models produce similar cross-diagonal slices but the arbitrary-inhomogeneity-based method provides better fit for diagonal slice as shown in Fig. 5(c).
The comparison of goodness of fit is given in Table \ref{tab:r2_comparison}.
\begin{table}[ht]
\centering
\caption{Comparison of $R^2$ values for Gaussian and Arbitrary inhomogeneity models}
\begin{tabular}{lccc}
\hline
\textbf{Model} & \textbf{Full 2D Spectrum} & \textbf{Diagonal Slice} & \textbf{Cross-Diagonal Slice} \\
\hline
Gaussian Inhomogeneity & 0.90 & 0.92 & 0.98 \\
Arbitrary Inhomogeneity & 0.95 & 0.9998 & 0.97 \\
\hline
\end{tabular}
\label{tab:r2_comparison}
\end{table}
The significantly higher values of $R^2$ for the whole 2D spectrum and diagonal slice quantify the improved fitting using arbitrary inhomogeneity method compared to Gaussian inhomogeneity.
The $R^2$ values for cross-diagonal slice are almost equal; these values are lower than those for the diagonal slice due to the presence of a weak biexciton peak around 1653 meV, which is not included in our model.

\section{Extensions of the model}
As discussed in the main text, we have used several assumptions to model the optical response of the two-level system.
These assumptions enable us to implement a computationally-efficient fitting procedure.
However, depending on the system under study, it might be necessary to perform these calculations beyond these assumptions.
We propose a few extensions to this model and discuss their consequence to the fitting procedure.

\subsection{Nonexponential decay dynamics}
In our simulation framework, we assume exponential decay for all oscillators, which is used in the optical Bloch equations (OBEs) for two-level systems.
However, this assumption may not hold in cases involving many-body interactions \cite{Kumar2024} or non-Markovian relaxation dynamics \cite{Lorenz2005}.
We can incorporate these phenomena by modyfying the homogeneous response function $S_h$ in Eq. (1) of the main text appropriately.
The anharmonic model for many-body interactions can be incorporated analytically \cite{Singh2016a}, without any other significant changes to the simulation.
However, numerical calculations are required for solving the modified OBEs \cite{Shacklette2003}.
For non-Markovian processes too, the nonlinear response can be calculated numerically after incorporating a frequency-frequency correlation function \cite{Mukamel1995}.

We would need to modify the fitting procedure in order to incorporate the above extensions to the model.
The most general route would require extracting the entire BSDF through the fitting procedure.
This situation can be significantly simplified by assuming $r_w = 1$, which is reasonable for $T \approx 0$.
Still, Eq. (6) of the main text, won't be valid.
Thus, the entire oscillator distribution will need to be defined by an array, that would be kept as a free parameter in order to calculate the best-fit spectrum.
This global fitting procedure can be computationally demanding due to the complexity of the equations.
The numerical calculations, if required, will also increase the computational cost even further, which may likely result in the fitting procedure being untenable.

\subsection{Variation in homogeneous linewidth}
We have assumed a constant homogeneous linewidth across the entire inhomogeneous distribution. 
However, the homogeneous linewidth may vary across different oscillators \cite{Moody2011}.
To account for this, we can generalize Eq. (2) in the main text by introducing oscillator-dependent decay rates along both the $\tau$ and $t$ axes:
\begin{equation}
    S(\tau,T,t) = \sum_{i,j=1}^{N}S_{h}(\tau,T,t,\gamma^{[i]}_{\tau},\gamma^{[j]}_t,\Omega^{[i]}_{\tau},\Omega^{[j]}_{t}) g_{\text{2D}}(\Omega^{[i]}_{\tau},\Omega^{[j]}_{t},M)
    \label{eq:rephasing_2D_gamma}
\end{equation}
where $S_h$ is also written in generalized form. 
\begin{equation}
   S_{h}(\tau,T,t,\gamma_{\tau},\gamma_t,\Omega_{\tau},\Omega_{t}) \propto e^{i\left(\Omega_{\tau}\tau - \Omega_T T -\Omega_{t}t\right)} e^{-(\gamma_{\tau}\tau +\gamma_T T+\gamma_{t}t)}\Theta(\tau,T,t).
   \label{eq:rephasing_homo_gamma}
\end{equation}
Here, $\gamma^{[i]}_\tau$ and $\gamma^{[j]}_t$ are the values of homogeneous linewidth corresponding to delays $\tau$ and $t$ respectively for the coordinate $(\Omega^{[i]}_{\tau},\Omega^{[j]}_{t})$.
Once again, if we assume $r_w = 1$, we can significantly simplify \eqref{eq:rephasing_homo_gamma} by using $\gamma^{[i]}_{\tau}=\gamma^{[j]}_t=\gamma^{[j]}$ where $\gamma^{[j]}$ corresponds to the homogeneous linewidth of oscillator corresponds to $\Omega^{[j]}$. 
Hence, \eqref{eq:rephasing_2D_gamma} can be written as
\begin{equation}
    S(\tau,T,t) = \sum_{i,j=1}^{N}S_{h}(\tau,T,t,\gamma^{[j]},\gamma^{[j]},\Omega^{[i]}_{\tau},\Omega^{[j]}_{t}) g_{\text{2D}}(\Omega^{[i]}_{\tau},\Omega^{[j]}_{t},M)\delta_{ij},
    \label{eq:rephasing_2D_gamma1}
\end{equation}where the Kronecker $\delta$ function $\delta_{ij}$ is included to ensure $r_w=1$.
Eq. (6) in the main text should also be modified to
\begin{equation}
\label{eq:Diag_supp}
\tilde{S}_{\text{Diag}}(\omega_{\tau'}) = \sum_{i=1}^{N} g_{\text{1D}}(\omega^{[i]}_{\tau'}) \cdot \frac{1}{\left( \gamma^{[i]}_{01} \right)^2 + \left( \omega_{\tau'} - \omega^{[i]}_{\tau'} \right)^2 }.
\end{equation}
Once again, we need to keep the entire array of energy-dependent $\gamma_{01}$ as a free parameter, which will make the fitting procedure significantly more complicated.
However, it might be possible to parametrize the dephasing rate as $\gamma_{01}(\Omega)$, as in Ref. \cite{Spivey2007}; this can reduce the number of free parameters and make the fitting procedure more tractable.

\subsection{Correlation spectrum}
\begin{figure}[t]
\centering\includegraphics[width=\textwidth]{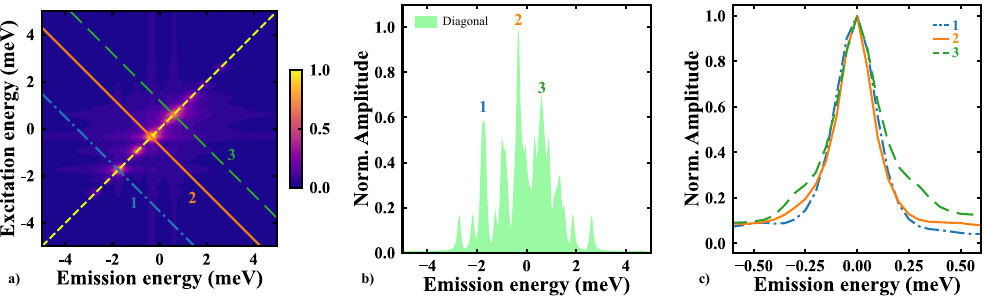}
\caption{(a)The simulated total 2D spectrum. (b)Diagonal slice of 2D spectrum. (c) comparison of cross-diagonal slices at different frequencies.}
\label{fig:corr2d}
\end{figure}
The simulation framework presented in this work can be extended to simulate 2D correlation spectrum for arbitrary inhomogeneity.
This capability would be useful for quantitative analysis of 2D spectra obtained from experiments performed in the pump-probe geometry \cite{DeFlores2007}, for instance, where the rephasing spectrum cannot be isolated.
For two-level system, the homogeneous response for 2D nonrephasing scan can be wriiten as:
\begin{equation}
   S^{NR}_{h}(\tau,T,t,\Omega_{\tau},\Omega_{t}) \propto e^{-i\left(\Omega_{\tau}\tau +\Omega_{t}t\right)} e^{-(\gamma_{\tau}\tau +\gamma_T T+\gamma_{t}t)}\Theta(\tau,T,t).
   \label{eq:non-rephasing_homo}
\end{equation}
The only change from the rephasing part is that the sign of phase evolution during delay $\tau$ is same as during delay $t$. 
Using the BSDF formalism, the nonrephasing third order nonlinear signal for an inhomgeneous distribution can be written as
\begin{equation}
    S^{NR}(\tau,T,t) = \sum_{i,j=1}^{N}S^{NR}_{h}(\tau,T,t,\Omega^{[i]}_{\tau},\Omega^{[j]}_{t}) g_{\text{2D}}(\Omega^{[i]}_{\tau},\Omega^{[j]}_{t},M)
    \label{eq:non-rephasing_inhomo}
\end{equation}
The third-order nonlinear signal for rephasing scan, $S^{R}(\tau,T,t)$ is given by Eq. (2) in the main text.
The superscript $R$ is used to distinguish from non-rephasing counterpart which is denoted by the superscript $NR$.
The total signal for 2D correlation spectrum is
\begin{equation}
    \tilde{S}(\omega_\tau,T,\omega_t) = \tilde{S}^R(|\omega_\tau|,T,\omega_t)+\tilde{S}^{NR}(\omega_\tau,T,\omega_t).
    \label{eq:2D}
\end{equation}
Figure \ref{fig:corr2d}(a) shows the 2D correlation spectrum for the oscillator distribution given in Figs. 2(a) and 2(b) of the main text.
The diagonal slice is shown in Fig. \ref{fig:corr2d}(b).
As expected, the peaks in the diagonal slice are narrower than those for the rephasing spectrum shown in Fig. 3(b) of the main text \cite{Khalil2003}.
Nevertheless, the crossdiagonal slices shown in Fig. \ref{fig:corr2d}(c) highlight the different crossdiagonal lineshapes, which is analogous to the results for rephasing spectrum discussed in the main text.
Here too, Eq. (6) of the main text is not valid and would lead to increased computational cost for the fitting procedure.
\bibliography{OL_arbinhomo}
